\documentclass[preprint,12pt]{article}

\usepackage{amssymb}

\usepackage{amsthm}
\newtheorem{theorem}{Theorem}

\usepackage{amsmath, amsfonts}
\usepackage{algorithm}
\usepackage{algpseudocode}
\usepackage{comment}
\usepackage{caption}
\usepackage{subcaption}
\usepackage{graphicx}

\def\Dart{\mathcal{D}_{art}}

\title{An Equity-Aware Recommender System for Curating Art Exhibits Based on Locally-Constrained Graph Matching}

\author{Anna Haensch
    \footnote{Data Intensive Studies Center, Tufts University, Medford, MA},
    Dina Deitsch
    \footnote{Tufts University Art Galleries, Tufts University, Medford, MA}}

\date{}
\begin{document}

\maketitle

\begin{abstract}
Public art shapes our shared spaces.  Public art should speak to community and context, and yet, recent work has demonstrated numerous instances of art in prominent institutions favoring outdated cultural norms and legacy communities. Motivated by this, we develop a novel recommender system to curate public art exhibits with built-in equity objectives and a local value-based allocation of constrained resources. We develop a cost matrix by drawing on Schelling’s model of segregation.  Using the cost matrix as an input, the scoring function is optimized via a projected gradient descent to obtain a soft assignment matrix. Our optimization program allocates artwork to public spaces in a way that de-prioritizes “in-group” preferences, by satisfying minimum representation and exposure criteria. We draw on existing literature to develop a fairness metric for our algorithmic output, and we assess the effectiveness of our approach and discuss its potential pitfalls from both a curatorial and equity standpoint.
\end{abstract}


\section{Introduction}\label{sec:introduction}

The placement of art in public spaces can have a significant impact on 
the individuals who view it. In the urban landscape, public art plays 
an important role in establishing the inclusionary/exclusionary role of 
urban regeneration, signaling whose interests and culture are being 
favored \cite{Sharp2005}.  An academic campus acts as a kind of city 
with its own physical configuration and established hierarchies, and 
the design of these physical spaces can deeply inform one's sense of 
belonging \cite{carvalho2018}.  Historically, most of the campuses 
across the United States were designed to support a majority white, 
male academic community.  More recently, there has been a movement 
towards creating more inclusive campuses, which call for a broad 
re-imagining of the spaces of campus and the people they serve 
\cite{Slee2011}.  As part of this, many college campuses have been 
turning to the visual arts as a means to address representation, using 
campus and public art collections to assist in dismantling the 
exclusionary spaces on campus.  Similarly, cultural institutions have begun to think about the ways that data science and AI can help to bring art collections in line with present day priorities \cite{Greenwald2021,Topaz2019,nytimes}. With that in mind, we propose a recommender system that curates exhibits in multiple spaces across an institutional campus with built in features that attempt to amplify underrepresented voices and curate culturally relevant exhibits.  We call this system {\em OpArt: The Optimal Art Curation Tool}.

\begin{figure}
\centering
\begin{subfigure}{.45\textwidth}
  \centering
  \includegraphics[width=.8\linewidth]{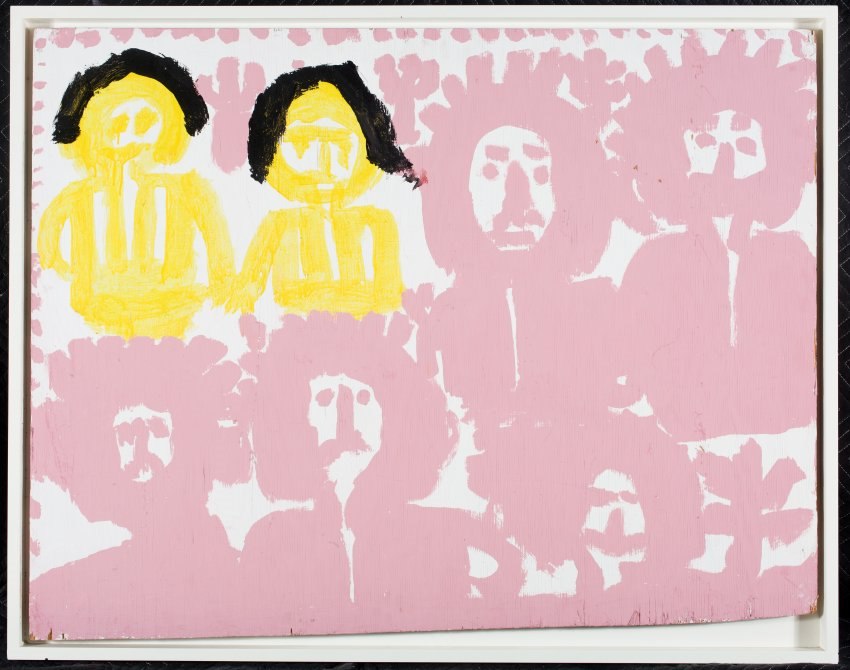}
  \caption{{\em Five Mauve Heads, Two }\\{\em Yellow Heads}, Mary Tillman \\Smith, 1989}
  \label{fig:smith}
\end{subfigure}%
\begin{subfigure}{.45\textwidth}
  \centering
  \includegraphics[width=.8\linewidth]{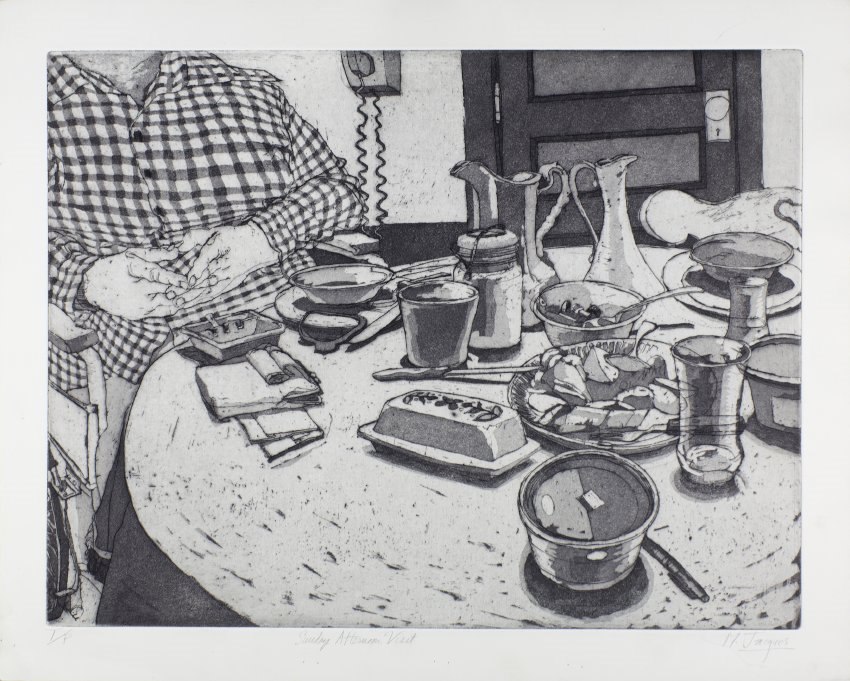}
  \caption{{\em Sunday Afternoon}, Michael L. Jacques, 20th Cent.\\}
  \label{fig:jacques}
\end{subfigure}
\caption{Images courtesy of Tufts University Art Galleries.}
\label{fig:collection}
\end{figure}

\subsection{Recommender Systems}\label{sec:rec_sys}

\begin{figure}[h]
    \centering
    \includegraphics[width = \textwidth]{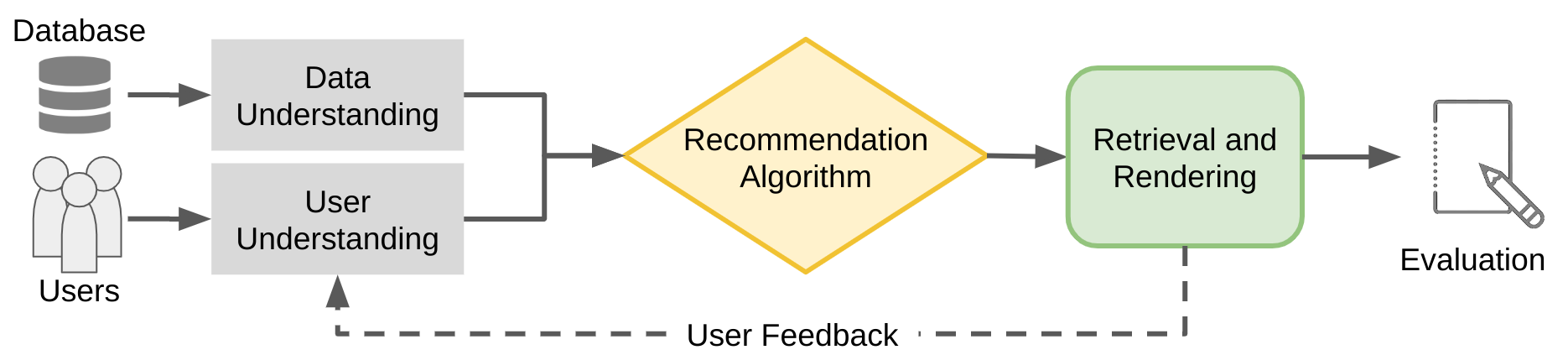}
    \caption{The pipeline for a recommender system follows the typical information retrieval pipeline.}
    \label{fig:pipeline}
\end{figure}

A recommender system is one form of information retrieval system \cite{ricci2021recommender}. The goal of this type of system is to present users with the items that most satisfy their needs, either implicit and explicit.  The pipeline for the recommender system we propose here follows the typical information retrieval pipeline as articulated in \cite{ekstrand2022fairness} (see Figure \ref{fig:pipeline}).  Underlying this system there is a database, $\mathcal D$, consisting of items, $d\in \mathcal D$.  Throughout this manuscript, we use as a running example the Tufts University Art Galleries Permanent Collection which exists in a database maintained using the proprietary industry-standard PastPerfect Museum Software \cite{art_database}.  To avoid any confusion, we will refer to this specific database as $\Dart$. 

A first step in the construction of a recommmender system involves data and item understanding.  Formally, this is established by a content representation of each item, and a metadata representation of the item, $\phi(d)$, which is composed of additional static data relevant to item $d$ such as creator details. Metadata representations are typically expressed as a vector quantity.  For example, two items represented in $\Dart$ are the paintings shown in Figure \ref{fig:collection}.  For these items, the content representations are the pixels comprising the digital renderings of the works themselves, and the metadata representations are the vectors consisting of static item attributes such as the ones shown in Table \ref{tab:metadata}.  An important piece of data understanding for most recommender systems -- but for the one proposed here in particular -- is the social and historical context of the creation of each item which help to identify the stakeholders in the items' usage \cite{becker2008art}.  This need for context must also be balanced with the knowledge that data limitations exist, and that treating identity parameters such as race and gender as a fixed and categorical attribute is reductive and can be at odds with efforts towards fairness and equity \cite{hanna2020towards}.

\begin{table}[]
    \centering
    \begin{tabular}{r|c|c|c|c}
                 & Artist Year of Birth & Artist Gender & Artist Race & $\cdots$\\
                 \hline 
                 \hline
        Figure 1 &  1904 & Woman & Black&\\
        Figure 2 &  1945 & Man & White&\\
        \hline
    \end{tabular}
    \caption{Partial metadata representations, $\phi(\cdot)$, of two sample database items.}
    \label{tab:metadata}
\end{table}

Another important initial step in constructing the system is to identify a set of user needs, $Q$, which are typically called user ``queries" in information retrieval. In the present work we will consider individual user needs, $q\in Q$, at a local level (i.e. an individual user's query is represented differently depending on when and where it is made).  We will express a query in terms of its local feature representation, which we will write as $\rho(q)$ for each $q\in Q$. To turn to our running example, a user might might submit a query, $q$, asking to see an item by an artist who shares their own race and gender.  If the user identifies as a Black woman, then the query representation is given by $\rho(q) = (\text{Woman},\text{Black})$.  At this point we would be remiss not to point to the ongoing research of the scholars in algorithmic fairness who argues that it is not only reductive to treat identity in this fixed and vectorized manner \cite{hanna2020towards}, but it amplifies the ``hierarchical logic" that produces these limiting categories to begin with \cite{hoffmann2019fairness}.  With this in mind, we will proceed with caution, and will continue to address some of these shortcomings as they arise throughout.

At the heart of this system is the recommendation algorithm of which a critical component is the scoring function, $s(q,d)$, which places a utility score on item $d$ for query $q$ indicating the extent to which $d$ satisfies the query.  There are many classes of algorithms commonly used at this step and for more details on these, the reader is direct to \cite{schutze2008introduction}.  In this paper we focus on a specific algorithm derived from graph matching; such algorithms have been used successfully for data mining and retrieval in the cases cases where syntax and data context are relevant to scoring \cite{brugger2008generalized, riesen2021novel,bislimovska2011content}.  After the algorithm has been executed, top scoring items are retrieved.  Keeping with our example of the user who requested a specific piece of art: at this step they might be shown Figure \ref{fig:smith}.  Once this is done, user feedback can be reintroduced into the system to refine user understanding. A final and essential part of this work, is the evaluation step.  The goal of the evaluative step is to establish the extent to which the algorithmic output satisfies individual user queries, or in the case of group recommendations, the goal is to determine whether all members of the group were treated fairly \cite{kaya2020ensuring}. 

\subsection{The OpArt Recommender System}\label{sec:op_art}

\begin{figure}
    \centering
    \includegraphics[width = \textwidth]{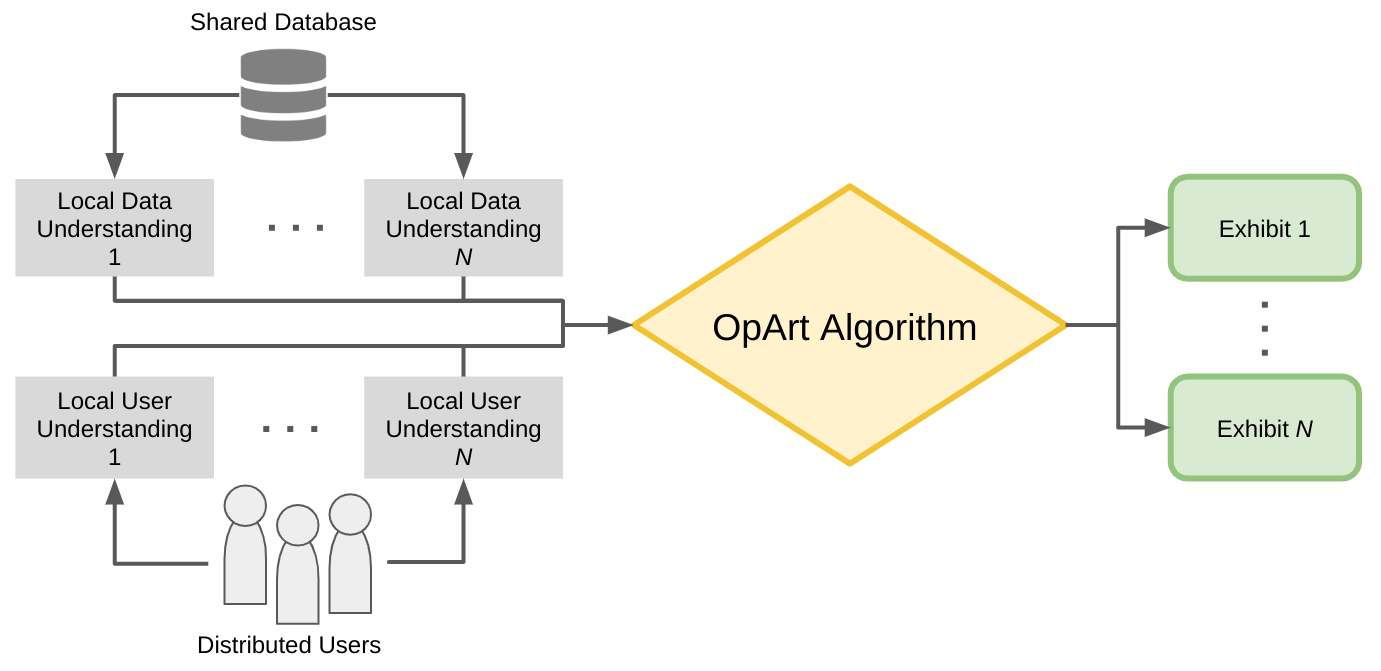}
    \caption{The OpArt pipeline identifies local user needs and relies on local data understanding to curate multiple exhibits with a shared set of resources.}
    \label{fig:OpArt_pipeline}
\end{figure}

The goal of the OpArt recommender system is threefold: first, to provide recommendations that are relevant to individual level queries; second, to provide recommendations that, in aggregate, are relevant at the global level; and third, to do all of this with a shared and limited set of resources. Unlike traditional recommender systems that algorithmically curate items for the individual user, OpArt is curating public art displays that need to satisfy many users concurrently (see Figure \ref{fig:OpArt_pipeline}.  Therefore, we will be interested in the score, $s$, described above, but extrapolated to a large set of users, the precise formulation of which will be given in Section \ref{sec:algorithm}. Built into this aggregate score will be the ability to give preferences to certain queries.  Specifically, queries made by locally underrepresented users will be given preference, and queries that seek out historically underrepresented artists will be given preference. 

\begin{figure}
    \centering
    \includegraphics[width = \textwidth]{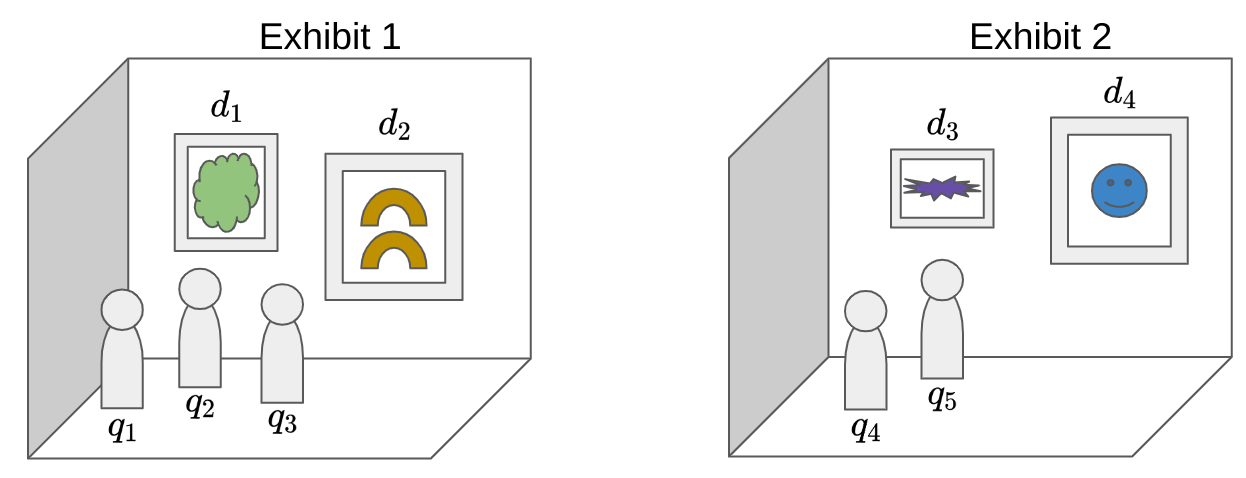}
    \caption{The goal of OpArt is to curate multiple physical art exhibits in spaces across an institution.}
    \label{fig:exhibits}
\end{figure}

After the algorithm is run, and recommendations are produced, the retrieval and rendering step of OpArt is carried out in the curation of multiple physical public art exhibits.  This is another feature that differentiates OpArt from typical information retrieval systems, namely, there is a natural capacity that exists for each item since one physical object can't appear in two places at once.  This means that sometimes multiple queries are at odds with one another.  For example, item $d_4$ from Figure \ref{fig:exhibits} might be the highest scoring item for both groups of queries, but it can only be shown in one exhibit space.  Consequently choices need to be made about which exhibit space and which queries are more highly valued.

While the question of value can be framed in multiple ways, 
through this work we hope to expand upon notions of value in art as a 
reflection of community identity, and as a mechanism for rethinking 
community belonging.  In particular, we explore one particular notion of value, 
motivated by the work of economist Thomas Schelling.  Schelling's model 
of segregation is an agent based model which demonstrates how individual 
preference for neighbors of the same race, even when mild, leads to 
segregation \cite{Schelling2006}. Schelling's work shows that in order 
to desegregate neighborhoods it is necessary to not only be tolerant of 
people different from oneself, but actively seek out diverse neighbors.  
Applying these principles to representation in public art, we develop 
our model of hanging art with a strong preference {\em against} ``in group" 
representation. 

The notation introduce throughout this section is summarized in Table \ref{tab:rs_notation}. The rest of the paper will be organized as follows.  Section \ref{sec:methodology} will cover the methodology proposed in this manuscript: in \ref{sec:representation} the database, users and associated representations will be formalized; in \ref{sec:algorithm} the recommendation algorithm will be derived with explicit values given for the scoring function.  Section \ref{sec:experiments} will report on experimental results, evaluating OpArt in terms of a group fairness metric. Section \ref{sec:discussion} will consist of a discussion and conclusions to be drawn from this work, as well as limitations of the methods presented here.   All of the tools and and data for this project are publicly available on Github \cite{opart_github}.  

\begin{table}[]
    \centering
    \begin{tabular}{c l}
         $d\in \mathcal D$& Item in a database\\
         $\Dart$ & Database of art collection \cite{art_database}\\
         $\phi(d)$& Metadata representation of item $d$\\
         $q\in Q$ & User need in space of all possible queries\\
         $\rho(q)$ & Feature representation of query\\
         $s(q,d)$ & Score of item $d$ relative to query $q$
    \end{tabular}
    \caption{Summary of notation introduced in Section \ref{sec:introduction}}
    \label{tab:rs_notation}
\end{table}

\section{Methodology}\label{sec:methodology}

Formally we approach this as a linear assignment problem with local constraints.  
First we consider exhibit locations as nodes on a graph, and 
consider the works of art in the collection as nodes on another graph.  
Classically, a graph matching problem would endeavor to match the art 
graph with the location graph in a way that preserves adjacencies within 
the graphs respectively.  In our approach, we also care about the local 
characteristics of the nodes, and therefore create a matching that 
preserves both inter- and intra- graph notions of closeness.  This will 
all be formalized in the following sections.  While much of what we present here can be abstracted to settings beyond OpArt, favoring technical abstraction over social context can often work at odds with goals of equity and fairness, and therefore we choose to present this as a context specific framework \cite{selbst2019fairness}. 

\subsection{Item and User Representation}\label{sec:representation}

The goal is to curate exhibits in $N$ different spaces using works from $\Dart$.  Each space can accommodate a different number of items and we denote this exhibition capacity, $h_n$.  If we fix an enumeration of spaces (e.g. by arranging them alphabetically), we express the capacity of the combined exhibits as a vector ${\bf h} := (h_1,..., h_n)$ and 
\begin{eqnarray*}
\sum_{n=1}^N h_n & = & \text{total number of items needed}.
\end{eqnarray*}
The artworks most relevant to the $n^{th}$ location will be determined by the combined needs of the users of that location.  In this work, we treat these needs as implicit, that is, we are assuming the needs of individuals rather than responding to direct requests which is in line with the common understanding of public art as tied to community and context \cite{cartiere2008practice}.  

Let $Q_n$ be the set of queries in location $n$.  Each of these local queries, $q\in Q_n$, has a vector representation as $\rho_n(q)$.  We will focus these queries on gender and race, supposing that each user's needs reflect their own race and gender which is consistent with the goal of public art to communicating belonging.  In this way we can think of $\rho_n(q)$ as embedded in a 2-dimensional space.  Since queries can exist with multiplicity, we defined $Q_n^{\text{mode}}$ as the most common query in $Q_n$. Recall that one of the explicit goals articulated in section \ref{sec:op_art} was to amplify queries from locally underrepresented users.  To do this, we will first compute a quantized version of the representation vector that give the proportional representation of each query,  
\begin{eqnarray*}
    \overline{\rho}_n(q)_j &:=& \frac{\mid \left\{q'\in Q_n: \rho_n(q)_j = \rho_n(q')_j\right\}\mid }{\mid Q_n\mid}. 
\end{eqnarray*}
Using this, we define a value function $v_n$ on the space of queries in $q\in Q_n$, given by 
\begin{eqnarray}\label{eq:v}
    v_n(q) & := & \sqrt{\sum_{j=1}^2\left[\overline{\rho}_n(q)_j - \overline{\rho}_n(Q_n^{\text{mode}})_j\right]^2}
\end{eqnarray}
which returns a small value for queries that are ``close" to the mode in terms of multiplicity, and a large value for rare queries.  Notice that this value function defines a Euclidean distance between the quantized representation vectors.

Since we are dealing with a physical curation process, there are only a limited number of items that can perfectly satisfy a query.  To make this concrete, the total number of works in $\Dart$ sorted by inferred artist gender and race, are give in Table \ref{tab:capacity}.  Fixing an enumeration of the entries using e.g. lexicographic ordering, We can partition our database as $\Dart = \cup_{m=1}^M \mathcal{D}_m$.  Further, taking $k_m = \mid \mathcal{D}_m\mid $, we obtain a built-in capacity for every possible query, which we express in vector form as ${\bf k} : = (k_1,...,k_M)$. The choice of how to enumerate these categories doesn't matter, all that matters is that once chosen, they are fixed.  In this way, 
\begin{eqnarray*}
\sum_{m=1}^M k_m & = & \text{total number of items available}.
\end{eqnarray*}
Recall that a second explicit goal of OpArt is to boost queries for historically underrepresented artists.  Treating the items $\Dart$ as embedded in a 2-dimensional space, we obtain quantized metadata representations for each $d \in \Dart$, defined by 
\begin{eqnarray*}
    \overline{\phi}(d)_j & := & \frac{\mid \left\{ d'\in \Dart : \phi(d)_j = \phi(d')_j\right\}\mid}{\mid \Dart\mid }.
\end{eqnarray*}
It is easy to show that $\overline{\phi}(d)_j = \overline{\phi}(d')_j$ for all $d,d'\in \mathcal{D}_m$ for any $m$ and $j$, therefore we will abuse notation slightly and use $\overline{\phi}(m)$ to mean the image under $\overline{\phi}$ of any item in $\mathcal{D}_m$ when the context is clear.  Since it will be useful as a normalizing term, we will defined the component-wise product of this quantized vector as 
\begin{eqnarray*}
    \overline{\Phi}(m) :=  \overline{\phi}(m)_1 \cdot \overline{\phi}(m)_2.
\end{eqnarray*}
When $d$ is a work that is underrepresented in the collection then this value is smaller.  Using this we define a second value function measuring the ``closeness" of a query to an item $d$ and to historical collecting patterns, 
\begin{eqnarray}\label{eq:w}
    w_{nm}(q) = \overline{\Phi}(m)\cdot \sqrt{\sum_{j=1}^2\delta_{\rho_n(q)_j\phi(m)_j}\cdot \left[\overline{\rho}_n(q)_j - \overline{\phi}(m)_j\right]^2},
\end{eqnarray}
where $\delta$ is the usual Kronecker delta function.  If the representations of $q$ and items in $\mathcal{D}_m$ are different in every component, then this value is 0.  In the case that they agree on a component, then the size of determined by their respective proportional shares in $Q_n$ and $\Dart$.  All of this can be easily extrapolated to a higher dimensional space. The notation introduced in this section is summarized in Table \ref{tab:methods_notation}.

\begin{table}[]
    \centering
    \begin{tabular}{c l}
        ${\bf h} = (h_1,...,h_N)$ & Vector of $N$ exhibit capacities\\
        ${\bf k} = (k_1,...,k_m)$ & Vector of $M$ query group capacities\\
        $Q_n$ & Set of queries in location $n$\\
        $Q_n^{\text{mode}}$ & The most common query in $Q_n$\\
        $\rho_n(q)$ & Vector representation of query $q$ at location $n$\\
        $\overline{\rho}_n(q)_j$ & Proportion of queries in $Q_n$ with vector representation $\rho_n(q)_j$ \\
        & in the $j^{th}$ coordinate\\
        $\overline{\phi}(d)_j$ & Proportion of work with metadata representation $\phi(d)_j$ \\
        & in the $j^{th}$ coordinate\\
        $v_n(q)$ & Closeness of query $q$ to the location mode in $Q_n$\\
        $w_{nm}(q)$ & Closeness of a query $q$ to item $d\in \mathcal{D}_m$ and historical collecting patterns\\
        $c(n,m)$ & Cost of including works from group $m$ in an exhibit at location $n$\\
        $S(n,m)$ & Overall score of group $m$ items for location $n$ exhibit.
    \end{tabular}
    \caption{Summary of notation introduced in Section \ref{sec:methodology}}
    \label{tab:methods_notation}
\end{table}

\subsection{Recommendation Algorithm}\label{sec:algorithm}

Now we are ready to define the algorithm at the heart of this system.  Recall that we are interested in an aggregate score for each item and location. To arrive at this score, we will first compute a cost function, $c$, which gives the cost for showing an item from $\mathcal{D}_m$ in location $n$, 
\begin{eqnarray}
    c(n,m) = \frac{\exp\left\{\sum_{q\in Q_n} \frac{-\alpha \cdot v_n(q)}{\beta \cdot w_{nm}(q)}\right\}}{\sum_{m'=1}^M\exp\left\{\sum_{q\in Q_n}\frac{-\alpha \cdot v_n(q)}{\beta\cdot w_{nm'}(q)}\right\}}.
\end{eqnarray}
where $\alpha$ and $\beta$ are positive parameters to be tuned. This cost function is derived from the standard multinomial logistic regression model, where $c(n,m)$ is taken as the probability of items from $\mathcal{D}_m$ given location $n$.  This choice of cost function is motivated by the observation that the value of particular item group to the query is exponentially weighted by considering the extent to which the query is representative of the location. The values of this cost function can be stored in an $N\times M$ matrix, $C$, where the entry in row $n$ and column $m$ is $c(n,m)$, the cost of showing items from $\mathcal{D}_m$ in location $n$. 

With this in mind, the aggregate score function described in section \ref{sec:op_art} can be written as an $N\times M$ matrix, $S$ where the $nm^{th}$ entry is the score of items from the $m^{th}$ group in the $n^{th}$ location.  The optimal score function is obtained by minimizing the cost while adhering to the capacity constraints, ${\bf h}$ and ${\bf k}$.  The initial proposed optimization program is 
\begin{eqnarray}
    \min_{\mathbb S}\quad \text{trace}(C^\intercal S)
\end{eqnarray}
where $\mathbb S := \{S\in \mathbb{R}^{N\times M}, S\succeq 0, S{\bf 1} = {\bf h}\}$ and ${\bf 1}$ denotes the all-ones column vector (i.e. $S{\bf 1}$ denotes the $m$-fold sum taken over the columns of $S$). To avoid the case where a given object is overly allocated, we further limit the number of times an object group can be utilized.  We do this via a penalty term that discourages objects from the $m^{th}$ group to be assigned to more than $k_m$ locations.  With this, the full optimization program is given by 
\begin{eqnarray}\label{eq:opt}
    \min_{\mathbb S} \quad \text{trace}(C^\intercal S) + \frac{\lambda}{2}\|S^\intercal {\bf 1} - {\bf k} \|_F^2
\end{eqnarray}
where $\lambda$ is a regularization parameter to be tuned.  Here we explicitly set ${\bf h}$ feasibility constraints (i.e. the locations requesting exhibits are fixed) but impose no such constraints for ${\bf k}$, since we would like to leave open the option for new artwork acquisition.  This means that all assignments are guaranteed to satisfy the space constraints, although some will deviated from art-in-hand, which raises interesting questions from a curatorial standpoint. The optimization program in equation \ref{eq:opt} is convex and we solve it using a projected gradient descent scheme that alternates between taking a gradient descent step and projecting onto the set $\mathbb S$.  This algorithm originates in \cite{condat2016fast}, but for clarity or exposition we include it here as Algorithm 1.  Next, we summarize several important aspects of the algorithms: initialization,  convergence, step size and stopping criterion.

\begin{algorithm}
\caption{A projected gradient descent algorithm to solve \eqref{eq:opt}}
\begin{algorithmic}[1]
\Statex
\State \textbf{Initialization:} Set $S^{(0)}$, $\lambda$, $\mathbf{k}$, $\mathbf{h}$, $\epsilon$ (step size), maxiterations (maximum number of iterations)
\For {i = 1:maxiterations}
\State Gradient descent: $\bar{S}^{(i)} =  S^{(i-1)}- \epsilon(C)-\epsilon\lambda\mathbf{1}(\mathbf{1}^\intercal S^{(i-1)}-\mathbf{k}^\intercal)$.
\State Project $\bar{S}^{(i)}$ onto the set $\mathbb S$ and obtain $S^{(i)}$ as follows: 
\For {n= 1:N}
\State Sort the entries of  row $n$ of $\bar{S}$ in decreasing order: $u_1\ge u_2\ge ... \ge u_M$. 
\State Define $L$: $L = \underset{1\le k\le M}{\max}\,\{k\mid \frac{\sum_{r=1}^{k} u_k-h_n}{k}<u_k\}$
\State Define $\tau$: $\tau = \frac{\sum_{\ell = 1}^Lu_\ell - \mathbf{h}_n}{L}$
\State Update row $n$ of $\bar{S}^{(i)}$: 
$S^{(i)}(n,m) = \max\left(\bar{S}^{(i)}(n,m)-\tau,0\right)$ for all $m=1:M$
\EndFor
\EndFor
\end{algorithmic}
\end{algorithm}

\textbf{Initialization}:  Since \eqref{eq:opt} is a convex 
optimization program, all the minimizers, $S^{*}$, i.e. the optimal 
assignment matrices, have the same objective value. However, these 
global minimizers themselves could be different from each other. Given 
that, it is of interest to assess the role of initialization on what 
kind of global minimizer is attained. Towards quantifying this, we consider three initializations for all experiments discussed in the paper. The first is initializing the permutation based on the ground truth assignment (i.e. what is currently on display, denoted $S^{\text{current}}$). Another initialization we consider is uniform assignment where each art piece is equally likely to be assigned to each building. Finally, we initialize by sampling uniformly randomly from the set $\mathbb S$ using the algorithm in \cite{smith2004sampling}.

\textbf{Convergence and step size}: To choose the step size, we first compute the Lipschitz constant of the gradient of the objective function in \eqref{eq:opt}. This quantity is useful in establishing the convergence of gradient descent.  Let $F(S)$ denote the objective function in \eqref{eq:opt}. The gradient of $F$ is Lipschitz continuous with parameter $L>0$ if $\|\nabla F(S_1)-\nabla F(S_2)\|_F\le  L\|S_1-S_2\|_F$ for all assignment matrices $S_1, S_2$. It can be verified that the Lipschitz constant of the gradient of $F$ is $L = \lambda N$.  We use the following standard result on the convergence of  projected gradient descent \cite{bubeck2015convex}.
\begin{theorem} [\cite{bubeck2015convex} (Theorem 3.7)] 
The projected gradient descent algorithm with $\epsilon =\frac{1}{L}$ satisfies
\[
f(S^{(k)})-f(S^*) \le \frac{3\epsilon\|S^{(0)}-S^*\|_F^2+f(S^0)-f(S^*)}{k},
\]
where $S^{(0)}$ denotes the initial assignment matrix, $S^{*}$ denotes the optimal solution and $S^{(k)}$ denotes the estimate of the assignment matrix at the $k^{th}$ iteration. 
\end{theorem}
\noindent Given the above convergence result, for all our experiments, we set $\epsilon = \frac{1}{2N}$. 

\textbf{Stopping criterion}:  The stopping criterion is the maximum number of iterations set to $1000$ which for all experiments is consistent with convergence in objective value. 

\textbf{Complexity}: In each iteration, the dominant computations are the gradient descent step and the projection step. The gradient descent step costs $O(NM)$ per iteration. Each projection of a row of $S^{(k)}$ costs $O(M\log(M))$. Therefore, the per-iteration cost of projection is $O(NM\log(M))$. We see that for $O(1)$ iterations, the algorithm is efficient scaling linearly with number of locations and item groupings. 

\textbf{Thresholding}: The soft assignment matrix is converted to a hard assignment matrix by rounding entries to nearest integer values.

From a practical point of view, if the optimization algorithm obtains 
an assignment matrix that differs significantly from the current 
assignment, significant costs might be incurred to make the necessary changes. With that, it might be useful in certain cases to make a gradual change. Rather than seeking globally optimal solutions that might be hard to realize in a timely manner, the practitioner might be interested in improving the current assignment gradually. To reflect this in our model, we make use of the existing assignment matrix, denoted by $S^{\text{current}}$ and modify \eqref{eq:opt} as follows

\begin{eqnarray}\label{eq:opt_with_prior}
\underset{
\mathbb S
}{\min}\quad \text{trace}(C^\intercal S)+\frac{\lambda}{2}\| S^\intercal\mathbf{1}-\mathbf{k}\|_{F}^2 +\frac{\tau}{2}\| S -S^{\text{current}}\|_{F}^2.
\end{eqnarray}
Above, the last term is a regularization term that encourages the optimized assignment matrix to not be far away from the current assignment. From an optimization perspective, equation \eqref{eq:opt_with_prior} is strongly convex for which there is a unique globally optimal solution. By varying the parameter $\tau$, the practitioner will obtain assignment matrices that are close to the current assignment while also optimizing the assignment problem of matching  exhibit pieces to locations. Complete details of the optimization program are given in Appendix \ref{appendix}.  Important notation introduced in this section is summarized in Table \ref{tab:methods_notation}.

\section{Experiments}\label{sec:experiments}

\subsection{Data Preparation}
To test the OpArt system, we run several experiments on the Tufts University campus. The current public art on campus consists of exhibits across 23 buildings, these buildings make up the rows of $S^{\text{current}}$ defined in section \ref{sec:algorithm}. The number of works from each group, $\mathcal{D}_m$, in each of the buildings make up entries in each row.  Summing across the columns we obtain ${\bf k}$, which we can think of as the total number of picture hooks in place at each location.  

These locations consist of academic building (that house academic departments and classroom space), resident halls (that also house student services and meeting space), administrative buildings, and what we will refer to as public use buildings such as libraries, theaters, chapels, and dining areas. The present use of each of the 23 buildings of interest can be easily obtained by querying the Tufts University Mobile Map \cite{mobile_map}.  Specifically, we can learn which academic programs are served by each building.  This information is then merged with student enrollment data obtained from the Tufts University Enrollment Calculator \cite{enrollment_calculator}.

Our goal is to use OpArt to set up art exhibits in each of the buildings that serve the students in the building according to our scoring method. We can't -- and shouldn't -- know exactly which students are going where, but we can make a best guess by combining what we know about the academic departments in each building, and the demographics of each department.  With this in mind, for each student we generate a daily path through campus using the following rules:

\begin{enumerate}
\item Each student visits the building(s) housing their primary department of enrollment. 
\item 1\% of students at random are assigned to buildings that house administrative space.
\item 2\% of students at random are assigned to buildings that house public space.
\item Students are split among residence halls at random according to the known number of available beds.
\end{enumerate}
In choosing the percentages for steps 2 and 3 are are assuming that each administrative office serves roughly 1$\%$ of the student body each day, while public spaces serve roughly 2$\%$.  Although one could easily adjust these percentages. Since the process contains some randomness we carry it out many times and report on aggregate results.  

\begin{figure}
    \centering
    \includegraphics[width = .7\textwidth]{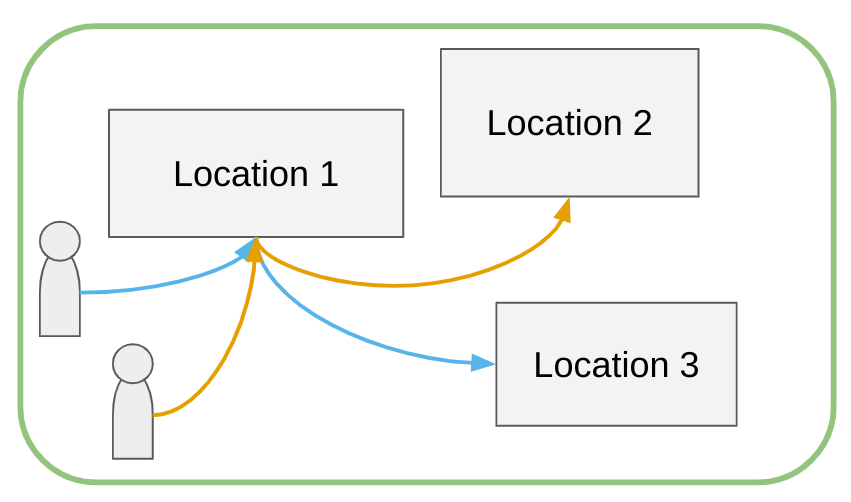}
    \caption{The daily path of a user is set deterministically by the user's primary department of enrollment and stochastically by the user's visit to public, administrative, and residence halls spaces.  Here two user paths are shown in blue and orange.}
    \label{fig:path}
\end{figure}

Here it is important to revisit the discussion from section \ref{sec:rec_sys} where we pointed out some of the limitations of categorical and static representations of racial and gender data.  In the institutional data that we've collected students are all classified biological sex terms ``Male" and ``Female," and we chose to modify this slightly in favor of the socialized gender terms ``Man" and ``Woman."  However, we also want to point out the erasure of non-binary students that results from this data gathering practice.  There are many students at our university who don't identify as ``Male" or ``Female," who are not given the opportunity to be counted.  We would also like to point out that the racial classifications used in our institutional data collection follows the racial groupings used by the US census prior to 2020, which have since been revised to allow for broader more intersectional racial and ethnic categories \cite{census}.  At present, we are working with the data we have, but we consider this a call to action to rethink institutional data collection and its goals and purposes.

Using this data, multiple experiments were carried out to curate exhibits across the 23 buildings.  All the algorithms in this paper are implemented in Python.  All necessary data, scripts, supporting Jupyter notebooks, and a README with instructions for recreating the results here are included in the OpArt Github repository \cite{opart_github}.  

\subsection{Results}\label{sec:results}

For the 
experiments, it is necessary to suitably set the hyper-parameters 
$\alpha$, $\beta$, $\lambda$ and $\tau$ that determine the optimal 
assignment matrix to be obtained from solving 
\eqref{eq:opt_with_prior}. For all our numerical experiments, 
$\alpha=-1$. For the experiments, we set $\beta$ to values in the range 
$10^{-1}$ to $10^{15}$. For a fixed value of $\beta$, before setting 
$\lambda$ and $\tau$, it is crucial to find the scaling of the terms in 
the objective \eqref{eq:opt_with_prior}. In particular, let 
$f_1(S) = \text{trace}(C^\intercal S)$, $f_2(S)=\|S^\intercal\mathbf{1}-\mathbf{k}\|_{F}^2$ 
and $f_3(S)= \|S-S^{\text{current}}\|_{F}^2$ denote the first three 
terms of the objective in \eqref{eq:opt_with_prior} respectively. 
For a fixed value of $\alpha$ and $\beta$, $f_1$ is determined. To 
determine appropriate scaling for $\lambda$ and $\tau$, we sample $r$ 
assignment matrices uniformly from the set $\mathbb{S}$ and compute 
$f_2$ and $f_3$ for each sampled matrix. We now define the scaling 
factors $\lambda_s$ and $\tau_s$ as follows: 
$\lambda_s =\frac{1}{r}\sum_{i=1}^{r} f_2(i)/f_1$ and 
$\tau_s = \frac{1}{r}\sum_{i=1}^{r} f_3(i)/f_1$.  In our experiments, we 
have used $r=50$. The scaling factors ensure that the three terms are 
comparable. With that, to set $\lambda$ and $\tau$, we proceed as 
$\lambda = \lambda_s\cdot \bar{\lambda}$ and $\tau = \tau_s \cdot \bar{\tau}$ 
where $\bar{\lambda}$ and $\bar{\tau}$ are multiplicative factors that 
are chosen from the range $[1,10000]$.

The goal of this work is to allocate objects to locations with the objective or prioritizing 
a self-representative allocation for individuals who are both historically underrepresented 
in terms of the object and the location.  To evaluate the success of this objective we will 
use a fairness metric similar to what is introduced in \cite{yao2017beyond}.  
We define $\mathcal G$ to be the disadvantaged group, and $\neg \mathcal G$ to be the 
advantaged group, so $\mathcal G\cup \neg \mathcal G$ is the set of all users. 

We define the fairness of our assignment as 
\begin{eqnarray*}
U = E_{\mathcal G}[y] - E_{\neg\mathcal G}[y]
\end{eqnarray*}
where $E_{\mathcal G}[y]$ as average number of self-representative objects seen by a person in group $\mathcal G$,
\begin{eqnarray}\label{expectation}
E_{\mathcal G}[y] &=& \frac{1}{\mid {\mathcal G}\mid }\sum_{g\in {\mathcal G}}r(g)
\end{eqnarray}
where $r(g)$ is the expected number of representative objects seen by 
person $g$ which can be computing as an expected value conditioned on 
the buildings.  These results are shown for variable optimization 
methods in Table \ref{tab:gender}.  The primary difference between our 
fairness metric and that proposed in \cite{yao2017beyond} is that we 
don't take the absolute value since our present work is 
specifically focused on elevating underrepresented groups.

\begin{table*}[ht]
\caption{Mean $\pm$ standard deviation for $E_{\mathcal G}[y]$ for 
baseline (i.e. current) and optimized assignment where we consider 
$\mathcal G$ as the disadvantaged group by gender (i.e. "Non-Man") and 
race (i.e. "Non-White").  The first row shows the baseline, and each 
subsequent row shows an optimization with a different initialization 
procedure setting $S^0$ equal to a uniform matrix, $S^{\text{current}}$, or a random projection.  Statistics in this table 
are gathered over 50 re-samplings of user base.}
\label{tab:gender}
\centering
\begin{tabular}[t]{c|cc|cc}
& Man & Non-Man & White & Non-White \\
\hline
\hline
Baseline & 
$12.280 \pm 0.134$&
$2.456 \pm 0.021$& 
$13.309 \pm 0.129$& 
$1.591 \pm 0.016$\\
Uniform Init. & 
$10.203 \pm 0.111$&
$4.018 \pm 0.032$& 
$6.594 \pm 0.065$& 
$7.614 \pm 0.068$\\
Current Init. & 
$10.869 \pm 0.119$&
$3.375 \pm 0.044$& 
$7.780 \pm 0.107$& 
$6.552 \pm 0.059$\\
Random Proj. &
$10.357 \pm 0.273$&
$3.871 \pm 0.259$& 
$6.860 \pm 0.523$& 
$7.362 \pm 0.474$\\
\hline
\end{tabular}
\end{table*}

From Table \ref{tab:gender} we can clearly see that our proposed 
methods of optimization not only increase the expected average 
self-representative objects seen by individuals in the disadvantaged 
group, but also increase fairness, $U$, over the baseline. Moreover, the overall fairness of the optimization is not particularly 
sensitive to our choice of initialization and is relatively stable over 
multiple rounds of sampling, where each round of sampling can be 
considered as a different day or session. The first row of the 
table gives the baseline stats for the current assignment, and 
subsequent rows apply the different initialization procedures 
described in section \ref{sec:algorithm}.  Before carrying out 
simulations, we scale $\lambda$ (resp. $\tau$) by the order of magnitude of quotient 
of the first and second (resp. first and third) terms of \eqref{eq:opt_with_prior}
to ensure that the scales are comparable between terms.  We replace ${\lambda}$ and ${\tau}$ with the scaled coefficients, $\overline{\lambda}$ and $\overline{\tau}$.

\begin{figure*}
\centering
\includegraphics[width = 1\textwidth]{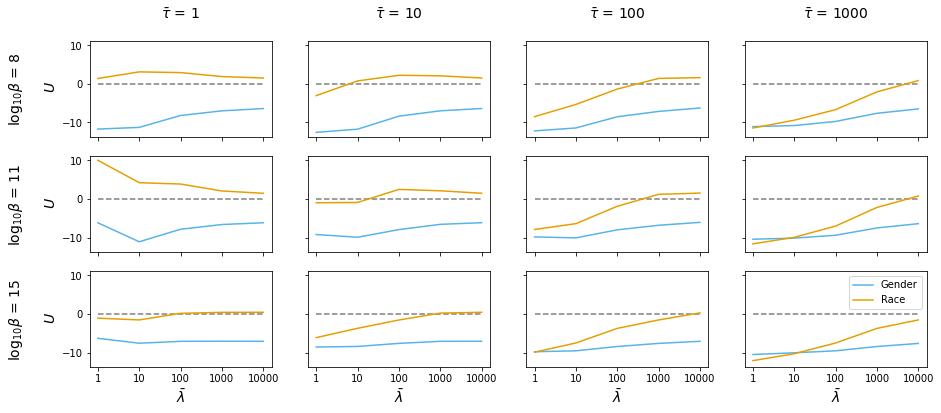}
\caption{Each plot in the grid above shows the $U$ score by demographic 
category.  The blue line indicates the score with $\mathcal{G}$ as the "non-male" group and the orange line indicates the score with $\mathcal{G}$ as the "non-white" group.  The dashed gray line at 0 is there to highlight when the assignment exhibits preferential treatment to the underrepresented group. }
\label{fig:lambda_beta_tau}
\end{figure*}

In Figure \ref{fig:lambda_beta_tau} we can see how the relative 
importance of model parameters are born out in the optimization 
algorithm. In particular, when $\lambda$ is small, the algorithm 
ignores artwork capacity constraints, which means that assignments 
might perform quite well in terms of fairness (see for example row 3, 
column 1 of Figure  \ref{fig:lambda_beta_tau} ), but are in-practice impossible, since they 
assign artworks beyond what the collection contains.  For example, such 
an assignment might place a high priority on artwork by Native Hawaiian 
Women across several buildings on campus, despite the fact that $\Dart$ holds no such works.  Perhaps this is a tool 
that could be used to inform new artwork acquisition.

In Figure \ref{fig:lambda_beta_tau}, we see that as $\tau$ 
increases the artwork hanging has a strong preference towards the 
status-quo.  In this case, disadvantaged groups continue to see very 
little art that reflects their own identity, until the $\lambda$ terms 
becomes large enough to overpower this preference for the status-quo 
(see for example row 1, column 4). From a curatorial standpoint, 
however, this slow movement towards more equitable art might be 
preferential, since acquiring and installing new exhibits takes both 
time and money.

\subsection{Group structure of optimal assignment matrices}

\begin{figure*}[h!]
\centering
\includegraphics[width = 1\textwidth]{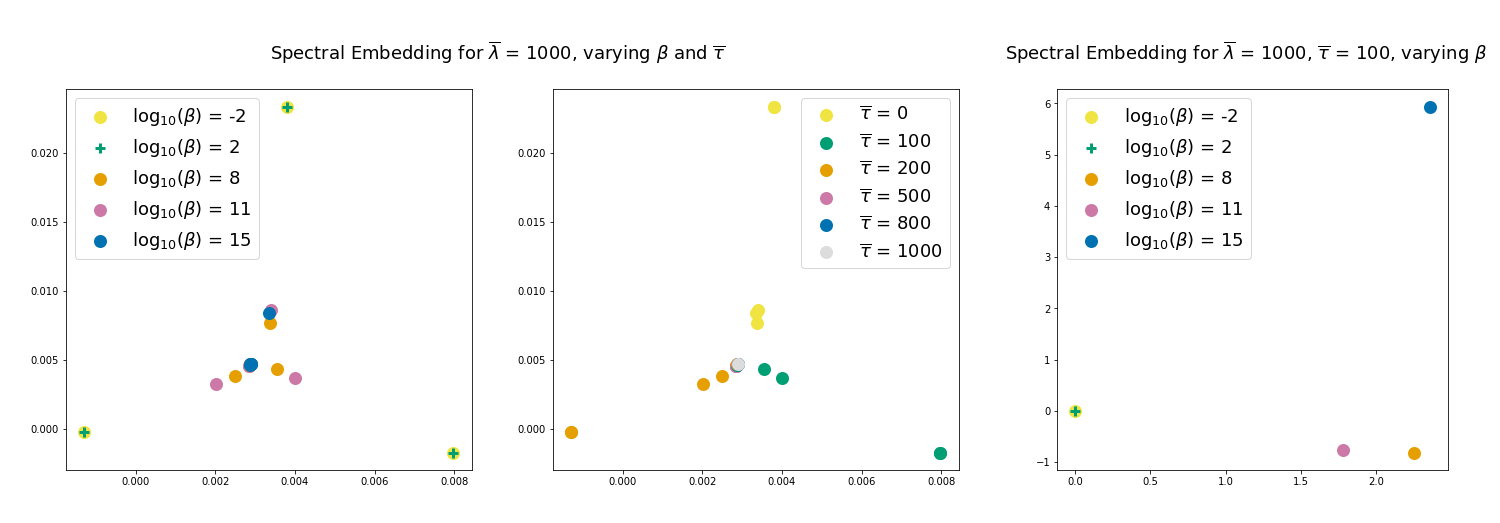}
\caption{Spectral embedding of optimal assignment matrices by varying the hyper-parameters: {\bf Left, Middle:} $\beta$ and $\overline{\tau}$ for fixed $\overline{\lambda}$, {\bf Right:} $\beta$ while $\overline{\tau}$ and $\overline{\lambda}$ is fixed.}
\label{fig:spectral_embedding}
\end{figure*}

As remarked in the previous section, the optimization program in \eqref{eq:opt} is convex
with all the minimizers sharing the same objective value. For a varying set of the hyper-parameters,
it is of interest to assess how ``different" the optimal permutation matrices are. To realize this goal,
we use the spectral embedding algorithm in \cite{ng2001spectral}. Spectral embedding takes
as an input a similarity matrix and outputs an embedding, in a certain dimension, in terms of point coordinates.
In our setting, the similarity matrix would inform us how two optimal assignment matrices are similar to each other.
First, under different choices of the hyper-parameters, we generate $r$ optimal score matrices. Let 
$W \in \mathbb{R}^{r\times r}$ denote the similarity matrix with $W_{i,j}$ denoting
the similarity between the $i^{th}$ optimal score matrix and the $j^{th}$ optimal score matrix. 
For our experiments, $W$ is defined as
\begin{eqnarray*}
    W_{i,j} &=& \begin{cases}
        \frac{1}{\|S_i-S_j\|_1} & \text{ if }i\neq j\\
        0 & \text{ if } i = j
    \end{cases}
\end{eqnarray*}
where $\|\cdot\|_1$ the entry-wise $\ell_1$ norm.  A larger entry would indicate that the two assignment matrices are
more similar. Given this, we run the spectral embedding algorithm and set the embedding dimension to $2$ for visualization
purposes. In Figure ~\ref{fig:spectral_embedding}, we show the spectral embeddings results. The first case concerns when $\bar{\lambda}$ is fixed and $\bar{\tau}$ and $\beta$ are varied. As $\bar{\lambda}$ is relatively larger, the art capacity constrains are met. We can clearly see that the different permutation matrices cluster based on the size of $\bar{\tau}$. In the right most plot, $\bar{\lambda}$ and $\bar{\tau}$ are fixed while $\bar{\beta}$ is varied. For this setting, we observe that there are different clusters corresponding to the different range of $\beta$. 

\section{Conclusion}\label{sec:discussion}

In this work we develop a recommendation system with build-in equity features.  The goal of the system is to curate multiple art exhibits across an institutional campus with the explicit goal of boosting minority group preferences while also prioritizing the work of underrepresented artists.  In this way, the work draws on Schelling's model which asserts the importance of out-of-group preferences in establishing diverse communities. These objectives are built into the system score through a cost function that establishes closeness between groups of queries and groups of items and optimizes score based on cost, thereby carrying out a version of locally constrained graph matching. 

On the one hand, this optimization tool is highly effective in highlighting the glaring blind spots in a collection and can indeed function as a guide for future acquisitions. In this specific case of $\Dart$, the history of collecting at the institution is a relatively short and passive one in that it has historically relied on gifts from alumni and would therefore tend to reflect those donors, who, following the historical demographics of the school, were and are majority white male.

On the other hand, curatorial work is multifaceted and the selection of 
artwork for campus spaces has to take into account a number of 
factors—from the basic logistics of security and direct sunlight 
(which will fade an artwork over time) to the content and subject 
matter of an artwork and its relevancy to the selected location. 
Similar to the methods used to leverage assignment priors in Section 
3.2, we could add a penalty term to leverage curatorial preferences by 
building and artwork. For instance, it might be poor judgment to hang a small, intimate photograph of a nude figure in a large dining hall where the space is large and loud and 1000s of students passed each day. The work would be subject to decay and fading from the light, possible damage from the proximity to the food, and the subject might not land well with a group of rowdy students after a long day, eating dinner.  Installing artwork in public spaces – be they a large lawn or library wall – requires a sensitivity to context that needs to run from the pragmatic to the conceptual – while yes, taking into account the maker of the artwork and its capacity to reflect our campus diversity as effectively as humanly possible. 

\section*{Acknowledgments}

The authors with to thank Abiy Tasissa for numerous discussion that significantly shaped the technical direction of this work. The authors also wish to thank the Data Intensive Studies Center at Tufts 
University for generous seed funding and Tufts University Art Galleries for inviting the authors to explore 
their collections and for granting access to their metadata.

 \bibliographystyle{elsarticle-num} 
 \bibliography{references.bib}
 
\appendix

\section{Artwork Capacities}

The following table gives artwork capacities for inferred race and gender as reported in \cite{art_database}.  It is important to note that these values are inferred and not self-reported.  The galleries has spent the last year reaching out to all of the known living artists to ask them to share their preferred identity details and look forward to having data that more accurately reflects the artist's lived realities.

\begin{table}[h]
    \centering
    \begin{tabular}{r|c|c|c|c|c|c}
            &White & Hispanic& Black & Asian & American Indian & Not \\
             && or Latinx & & & or Alaska Native & Inferred\\
             \hline
             \hline
        Man& 1425 & 131 & 54 & 36 & 19 & 2 \\
        Woman& 227 & 6 & 9 & 6 & 15 & 0 \\
        Not Inferred& 0 & 6 & 1 & 1 & 0 & 9 \\
        \hline
    \end{tabular}
    \caption{Artwork capacities based on inferred gender and race.}
    \label{tab:capacity}
\end{table}

\section{Details of optimization program}\label{appendix}

\begin{figure}[h!]
\centering
\includegraphics[width = 0.5\textwidth]{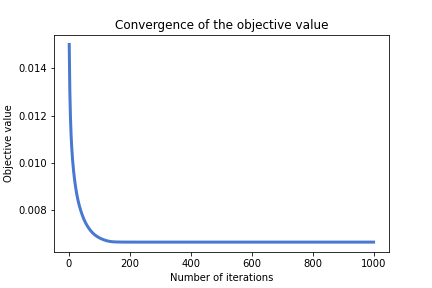}
\caption{Convergence of the projected gradient descent algorithm with $\beta= 100$, $\overline{\tau}=1$ and $\overline{\lambda}=1$}
\label{fig:convergence}
\end{figure}

We revisit the main optimization program.
\begin{equation}\label{eq:opt_with_prior_appendix}
\underset{
\mathbb S
}{\min}\quad \text{trace}(C^\intercal S)+\frac{\lambda}{2}\|S^\intercal\mathbf{1}-\mathbf{k}\|_{2}^2 +\frac{\tau}{2}\|S-S^{\text{current}}\|_{F}^2.
\end{equation}
Let $f_1(S) = \text{trace}(C^\intercal S)$, $f_2(S)=\|S^\intercal\mathbf{1}-\mathbf{k}\|_{F}^2$ and $f_3(S)= \|S-S^{\text{current}}\|_{F}^2$ denote the first three terms of the objective in \eqref{eq:opt_with_prior_appendix} respectively. The first term is linear in $S$ and hence convex. The second and third terms are composition of a convex function with affine functions  yielding convex functions. It follows that the objective in \eqref{eq:opt_with_prior_appendix} is convex. All in all, we have a convex program. As discussed in the main part of the paper, we employ projected gradient descent to optimize \eqref{eq:opt_with_prior_appendix}. From our numerical experiments, convergence in objective is attained before the maximum iterations set to $1000$. A typical instance of the convergence plot is shown in Figure~\ref{fig:convergence}.

To derive the step size for projected gradient descent, we rely on the Lipschitz constant of the gradient. Below, we detail the computation.  Hereafter, the objective in \eqref{eq:opt_with_prior_appendix} is denoted by $f(S)$. \\
\textbf{Computation of $\nabla f_1(S)$}: We compute $\nabla f_1(S)$ as follows.
\begin{align*}
\frac{\partial }{\partial S_{i,j}}  ( \text{trace}(C^\intercal S)) & = \frac{\partial}{\partial S_{i,j}}  \sum_{mn} C_{mn} S_{mn} \\                      & = \sum_{mn} C_{mn} S_{mn} \delta_{mi}\delta_{nj} \\
                & = C_{ij}		        
\end{align*}
Therefore, $\nabla f_1(S) = C$. 

\textbf{Computation of $\nabla f_2(S)$}: We compute $\nabla f_2(S)$ as follows.
\begin{eqnarray*}
\frac{\partial } {\partial S_{i,j}} \left (\|S^\intercal\mathbf{1}-\mathbf{k}\|_{2}^2\right) & = &\frac{\partial } {\partial S_{i,j}}  \left(\sum_{m=1}^{M} ( (S^\intercal\mathbf{1})_m-k_m)^2\right) \\
& = &\frac{\partial } {\partial S_{i,j}}  \left(\sum_{m=1}^{M} \left[\sum_{r=1}^{N} S_{rm}- k_m\right]^2\right) \\
& = &2\left( \sum_{r=1}^{N} S_{rj}-k_j\right) \\
&= &2(S^T\mathbf{1})_j-2k_j
\end{eqnarray*}
We note that the last term is independent of $j$. It follows that $\nabla f_2(S)= 2\mathbf{1}\mathbf{1}^\intercal S-\mathbf{1}\mathbf{k}^\intercal=2\mathbf{1}(\mathbf{1}^\intercal S-\mathbf{k}^\intercal)$.

\textbf{Computation of $\nabla f_3(S)$}: We compute $\nabla f_3(S)$ as follows.
\begin{eqnarray*}
\frac{\partial } {\partial S_{i,j}} \left ( \|S-S^{\text{current}}\|_{F}^2 \right) &= &\frac{\partial } {\partial S_{i,j}} \sum_{m,n}
\left(S_{m,n}-S^{\text{current}}_{m,n}\right)^2  \\
& = &2\sum_{m,n}  \left(S_{mn}-S^{\text{current}}_{mn}\right)\cdot \delta_{mi}\delta_{nj} \\
& = &2\left(S_{ij}-S^{\text{current}}_{ij}\right)
\end{eqnarray*}
Therefore, $\nabla f_3(S) = 2(S-S^{\text{current}})$.  We now proceed to compute the Lipschitz constant of the gradient as follows. For any $S_1, S_2$, 
\begin{eqnarray*}
\|\nabla f(S_1)-\nabla f(S_2) \| _F& =& \|(\lambda\mathbf{1}\mathbf{1}^\intercal-\tau\mathbf{I})(S_1-S_2)\|_F \\
& \le& \|\lambda \mathbf{1}\mathbf{1}^\intercal-\tau\mathbf{I}\|_2\, \|S_1-S_2\|_F
\end{eqnarray*}
The eigenvalues of $\lambda \mathbf{1}\mathbf{1}^\intercal$ are $0$ and $\lambda N$. It follows that  $\|\nabla f(P_1)-\nabla f(S_2) \| _F \le (N-\tau) \, \|S_1-S_2\|_F$. Hence, the Lipschitz constant of the gradient is $\lambda N-\tau$. 

\end{document}